\documentclass[conference,a4paper]{IEEEtran}%
\usepackage[utf8]{inputenc}
\usepackage{graphicx}
\usepackage{tabularx}
\usepackage{amsmath}
\usepackage{amssymb}
\usepackage{stackengine}
\usepackage{makecell}
\usepackage{multirow}
\usepackage{cite}
\usepackage{array}
\usepackage{bbding}
\usepackage{pifont}
\usepackage{wasysym}
\usepackage{comment}
\usepackage{caption}
\usepackage{subcaption}
\usepackage[noend]{algpseudocode}
\usepackage{color}
\usepackage{booktabs}
\usepackage{mathtools}
\usepackage{dirtytalk}
\usepackage{etoolbox}
\usepackage{tikz}
\usepackage{algorithm,algpseudocode}
\usepackage{url}
\usepackage{float}
\usepackage{stfloats}

\title{Assessing the Technical and Environmental Impacts of Energy Management Systems in Smart Ports}

\makeatletter
\newcommand{\linebreakand}{%
  \end{@IEEEauthorhalign}
  \hfill\mbox{}\par
  \mbox{}\hfill\begin{@IEEEauthorhalign} \vspace{-10cm}
}
\makeatother
\author{
  \IEEEauthorblockN{Youzhe Yang}
  \IEEEauthorblockA{
       \textit{LUT University,}\\
    Lappeenranta 53850, Finland \\
   }
  \and
  \IEEEauthorblockN{Hafiz Majid Hussain}
  \IEEEauthorblockA{
   \textit{LUT University,}\\ 
   Lappeenranta 53850, Finland  }
  \and
  \IEEEauthorblockN{Juha Haakana}
  \IEEEauthorblockA{
    \textit{LUT University,}\\
      Lappeenranta 53850, Finland \\
    } 
 \and
  \IEEEauthorblockN{Pedro Nardelli}
  \IEEEauthorblockA{
    \textit{LUT University,}\\
      Lappeenranta 53850, Finland \\
    } 
   
} 
\makeatletter
\patchcmd{\@maketitle}
  {\addvspace{0.5\baselineskip}\egroup}
  {\addvspace{-1\baselineskip}\egroup}
  {}
  {}
\makeatother

\begin{document}  

\maketitle
\begin{abstract}
A vital strategy for ports to mitigate the environmental impact of the maritime industry, while complying with frameworks such as the European Green Deal and the Sustainable Development Goals (SDGs), entails the systematic implementation of comprehensive energy management solutions. This paper provides a baseline evaluation of the energy management systems (EMSs) implementation and their impact on energy consumption, carbon emissions, and operational costs in smart ports. Initially, we provide a systematic review of the literature focusing on case studies from prominent ports, including Hamburg, Genoa, Jurong, and Shanghai Yangshan Phase IV.  The analysis emphasizes key aspects such as energy efficiency, reductions in emissions, and the minimization of operational costs. Subsequently, we formulate an optimization model to simulate load dispatch, carbon emission reduction, and transport scheduling. Results indicate that EMS deployment reduces annual energy consumption and carbon emissions significantly—by approximately 7–8\% and 11–12\% respectively—while achieving substantial cost savings 30\%. The study also identifies critical challenges, including system integration, data quality issues, cybersecurity risks, and the need for standardization. These findings provide valuable insights for port authorities and policymakers, supporting the transition toward more sustainable and efficient port operations.
\end{abstract}

\section{Introduction}
As the hubs of international logistics, ports are the gateways for the movement of goods and commodities across the world \cite{yau2020towards, hussain2020energy}. As globalization intensifies and the demand for maritime transport grows, ports are experiencing a substantial increase in cargo throughput and ship traffic \cite{belcore2023connected}. This growth, while economically beneficial, brings with it a range of environmental and technological challenges \cite{moffatt2025greenports}. One of the most pressing concerns is the significant rise in energy consumption and associated greenhouse gas emissions, particularly carbon dioxide, from port operations and associated logistics activities. Addressing these challenges necessitates a shift towards more sustainable and efficient port management strategies \cite{nguyen2022technical}.

Generally, ports have relied on conventional operational systems, which often lack the agility, coordination, and technological integration needed to minimize environmental impact \cite{pereira2025process}. These shortcoming results in energy inefficiencies, underutilization of resources, and a growing carbon footprint \cite{issa2023optimizing}. Therefore, it is imperative to adopt innovative solutions that not only streamline port operations but also contribute to environmental sustainability.

In response to these challenges, the concept of smart ports has emerged as a transformative paradigm. Smart ports integrate advanced technologies such as the Internet of Things (IoT), big data analytics, artificial intelligence, and cloud computing into port operations to enhance efficiency, safety, and sustainability \cite{ yang2018internet}. A central component of smart ports is the energy management system (EMS), which plays a pivotal role in monitoring, controlling, and optimizing energy consumption across port facilities \cite{clemente2023blue, hussain2022benchmarking}. EMS enables real-time data collection and analysis, facilitating informed decision-making and proactive energy-saving measures \cite{skok2024enhancing, hussain2024crossover, hussain2021packetized}. In addition, the implementation of EMS in smart ports not only improves operational efficiency but also significantly reduces energy waste and CO$_2$ emissions. By aligning energy demand with supply, automating processes, and integrating renewable energy sources, EMS contributes to the broader goal of environmental sustainability in maritime logistics \cite{chen2019constructing}.

In the framework of EMS within smart port environments, this paper conducts a thorough review of the integrated EMS practices utilized across four major ports: Hamburg, Genoa, Jurong, and Shanghai Yangshan. Our study provides a systematic comparison among these ports, highlighting the sophisticated capabilities of EMS, the key driving factors behind their implementation, as well as the core technologies and technical tools that are crucial for achieving sustainable and cost-effective operations in smart port contexts.

Furthermore, we develop an optimization model aimed at simulating load dispatch, carbon emission reduction, and transport scheduling. The results of these simulations illustrate a significant reduction in annual energy consumption and carbon emissions within the examined ports. The principal contributions of this research are delineated as follows:

\begin{itemize}\item A systematic analysis of EMS and its fundamental significance in the advancement of smart ports, specifically focusing on the four prominent ports: Hamburg, Genoa, Jurong, and Shanghai Yangshan.\item The development of an optimization model designed for load dispatch, carbon emission reduction, and transport scheduling.\item Conducting and discussing the results of the simulation outcomes.\end{itemize}

The rest of the paper is organized as follows. Section II
presents the integrated EMS in smart port- case studies review and proposed solution. Section III provides  the simulation results, and section IV concludes the paper.

\begin{table}[!t]
\renewcommand{\arraystretch}{1.3}
\centering
\caption{Energy Use and Emission Reduction Performance\cite{jiangsu_ecology_2024},  \cite{sciencedirect_2019}, \cite{ntu_2024}, \cite{port_of_hamburg_2024}, \cite{ruca_logistics_2024}}
\begin{tabular}{|p{2cm}|p{1.2cm}|p{1.2cm}|p{1.2cm}|p{1.2cm}|}
\hline
\textbf{Category} & \textbf{\centering Hamburg} & \textbf{\centering Genoa} & \textbf{\centering Jurong} & \textbf{\centering Shanghai} \\
\hline
Renewable Energy Share (2023) 
& 23.5\% 
& 40\% 
& 60\% 
& 48.8\% \\
\hline
Emission Reduction (2020--2023) 
& 38\% 
& 35\% 
& 15\% 
& 10\% \\
\hline
Energy Demand Scheduled (\%) 
& 20\% \newline (smartPORT App) 
& 22.4\% \newline (Port Community System) 
& 48.4\% \newline (JP Glass) 
& 30\% \newline (ITOS system) \\
\hline
\end{tabular}
\label{tab:energy_emission}
\vspace{-0.3cm}
\end{table}

\begin{table}[!t]
\renewcommand{\arraystretch}{1.3}
\centering
\caption{Technical and Automated Means}
\begin{tabular}{|p{2cm}|p{1.2cm}|p{1.2cm}|p{1.2cm}|p{1.2cm}|}
\hline
\textbf{Category} & \textbf{Hamburg} & \textbf{Genoa} & \textbf{Jurong} & \textbf{Shanghai} \\
\hline
Core Technologies 
& Digital twins\newline Predictive AI 
& AI analytics\newline Electrified AGVs 
& AI-driven SMES\newline 3D digital twin 
& AGV battery swap\newline Digital twin \\
\hline
Automation Impact 
& 20\% cargo flow\newline efficiency gain 
& 20\% faster\newline cargo clearance 
& 20\% turnaround\newline time reduction 
& 70\% labor cost\newline reduction \\
\hline
Cost Reduction via Smart Devices 
& 40\%\newline (AI optimization) 
& 30\%\newline (Digital twins \& BI) 
& 10\%\newline (SMES \& IoT) 
& 30\%\newline (Battery swap system) \\
\hline
\end{tabular}
\label{tab:tech_automation}
\vspace{-0.2cm}
\end{table}

\begin{table}[!t]
\renewcommand{\arraystretch}{1.3}
\centering
\caption{Energy Mix and Storage Patterns}
\begin{tabular}{|p{2cm}|p{1.2cm}|p{1.2cm}|p{1.2cm}|p{1.2cm}|}
\hline
\textbf{Category} & \textbf{Hamburg} & \textbf{Genoa} & \textbf{Jurong} & \textbf{Shanghai} \\
\hline
Key Renewable Sources 
& Shore power\newline Solar\newline Hydrogen 
& Solar\newline Hybrid grids 
& Solar\newline LNG\newline Thermal storage 
& Solar\newline LNG\newline Shore power \\
\hline
Energy Storage 
& Hydrogen\newline storage (planned) 
& Battery storage 
& Thermal storage 
& Battery storage \\
\hline
Grid Flexibility 
& Smart grids 
& Hybrid grids 
& IoT edge devices 
& Smart grids \\
\hline
\end{tabular}
\label{tab:energy_storage}
\vspace{-0.2cm}
\end{table}

\begin{table}[!t]
\renewcommand{\arraystretch}{1.3}
\centering
\caption{Alignment of Strategic Objectives and Policies}
\begin{tabular}{|p{2cm}|p{1.2cm}|p{1.2cm}|p{1.2cm}|p{1.2cm}|}
\hline
\textbf{Category} & \textbf{Hamburg} & \textbf{Genoa} & \textbf{Jurong} & \textbf{Shanghai} \\
\hline
Decarbonization Targets 
& Climate neutrality\newline by 2040 
& 35\% carbon\newline reduction by 2025 
& Fossil-free\newline by 2025 
& 50\% RE share\newline by 2030 \\
\hline
Policy Alignment 
& EU FuelEU\newline Maritime Directive 
& Italy’s PNRR\newline Recovery Plan 
& Singapore\newline Green Plan 2030 
& China’s\newline Dual Carbon Goals \\
\hline
\end{tabular}
\label{tab:policy_alignment}
\vspace{-0.2cm}
\end{table}


\section{Integrated EMS in Smart ports --Case studies review}

\subsubsection{Hamburg port}
A leading European logistics hub, integrates advanced EMS and port intelligence to enhance sustainability and operational efficiency. The EMS at Hamburg port is driven by the smartPORT energy strategy, emphasizing three pillars: renewable energy resources (RESs) integration, energy efficiency optimization, and green mobility transformation \cite{ruca_logistics_2024}. Among the three pillars, RESs initiatives include shore power, solar energy, and hydrogen storage. Recently, port was supplied 1,420~GWh of shore power, including  RESs, i.e.,  distributed solar panels (8.5~MW), which generated over 800~MWh annually. Likewise, the port authority aims to establish a hydrogen storage infrastructure to meet 70\% of Germany’s hydrogen import needs by 2030 \cite{hafen2025}. To accommodate seamless integration of RESs, numerous technological developments have been made, such as smartPORT framework predictive AI, smart grid technologies, and digital twins. Hamburg's smartPORT framework uses digital tools and data to optimize logistics and infrastructure. Over 300 road sensors monitor traffic and feed data into the smartPORT App for traffic coordination. The PortTraffic Center integrates transport modes, increasing cargo flow efficiency by 20\%.  .

\subsubsection{Genoa port}
Genoa port, a major Mediterranean hub, enhances sustainability through EMS, aligning with Italy’s National Recovery and Resilience Plan (PNRR) and EU goals. Its EMS focuses on solar energy, smart grid technologies, and vehicle electrification. The port integrates distributed solar, which produces 800 MWh/year to facilitate port operations and EVs, and as a result reduces 6,500 tons of CO$_2$ \cite{portgenoa2023}. Similarly, the implementation of a hybrid grid improves the balance between numerous RESs distribution, which further helps the port to reduce its environmental impact.  Additionally, the port leverages digital infrastructure, digital twins,  automation, and smart logistics to support energy monitoring, predictive maintenance, and operational cost reduction. Furthermore, the port community system (PCS) enables paperless stakeholder integration, accelerating cargo clearance while an IoT-based BI platform supports data analysis to integrate sea, rail, and road transport, boosting cargo flow by 20\% . 

\vspace{-0.1cm}
\subsubsection{Jurong Port}

Jurong Port in Singapore integrates EMS and port intelligence to drive efficiency and sustainability. RESs integration includes solar (9.5 MWp, 12 million kWh/year) and the smart multi-energy system (SMES), developed with Nanyang Technological University (NTU) and SP Group. SMES dynamically optimizes energy sources, saving 10\% energy and cutting carbon by 15\% \cite{disruptive2024}. Jurong’s JP Glass platform integrates operations with 3D visualization for berth monitoring and cargo optimization. Built on ArcGIS, it supports simulations for turnaround time improvement.  As mentioned by \cite {cash2024} that  7,000+ tons of annual CO$_2$ reduction, 57.4 moves/hour crane productivity, and 20,000+ TEUs throughput has achieved via SMES.Additionally, SMES cut energy costs by 10\%, and predictive maintenance reduced downtime by 40\%.

\subsubsection{Shanghai Yangshan Phase IV Port}

Shanghai Yangshan Phase IV, located in Shanghai, is recognized as the largest automated container terminal globally, exemplifying the integration of EMS and automation technologies. The terminal incorporates various RES, including 8.5 MW solar panels that generate approximately 8 million kWh annually, accounting for 30\% of the terminal's energy demand. Additionally, it features liquefied natural gas (LNG) bunkering and shore power systems that collectively supply 1,420 GWh per year. These initiatives have resulted in emission reductions exceeding 7,500 tons annually. The implementation of smart grid technologies facilitates the dynamic balancing of power demand and supply across multiple generation and consumption sources \cite{xinde_marine_news_2022}. Furthermore, the introduction of AGVs within the container management network has also contributed to a 40\% decrease in diesel fuel consumption \cite{stdc_shanghai_2025}.

A summary of the EMS adopted by the four ports is presented in the accompanying table \ref{tab:energy_emission}, \ref{tab:tech_automation}, \ref{tab:energy_storage},
\ref{tab:policy_alignment}.
. 

\subsection{Comparative Analysis: Renewables, Emissions, Costs, and Dispatch Efficiency}

his subsection provides a comparative analysis of the four ports concerning renewable energy shares, carbon emission reductions, cost savings, and dispatch efficiency \cite{jiangsu_ecology_2024, sciencedirect_2019, ntu_2024,port_of_hamburg_2024, ruca_logistics_2024}. Jurong Port leads in renewable energy usage, while Genoa Port and Shanghai Yangshan Phase IV utilize about 40\% and 50\% RES, respectively, with Hamburg Port showing the lowest share at roughly 23\%, all influenced by diverse governing policies. For emission reductions, Hamburg Port achieved the highest at 38\% through shore power and electric equipment, followed by Genoa Port at 35\% and Jurong Port at 15\% (constrained by its existing high efficiency); Shanghai Yangshan Phase IV recorded the lowest at 10\% due to infrastructure limitations. In terms of cost reduction, Hamburg Port again led with 40\% (driven by AI and hydrogen energy), while Genoa and Shanghai Yangshan Phase IV both achieved 30\% savings through digital tools and automation; Jurong Port saw a modest 10\% reduction due to its optimized structure. Finally, movement control efficiency highlights Jurong Port's lead at 48.4\% (enabled by its 3D Digital Twin and AI-optimized cargo sequencing), followed by Shanghai Yangshan Phase IV at 30\% (leveraging ITOS automation), Genoa Port at 22.4\% (aided by PCS and AI forecasting), and Hamburg Port at 20\% (constrained by fragmented data integration)

\subsection{SOLUTION PROPOSAL–EMS INTEGRATOR}

\subsubsection{Environmental Impact Modelling}

As mentioned earlier, one of the key roles of EMS is to reduce carbon emissions mainly by optimising energy consumption and introducing cleaner energy sources. The total carbon emissions can be expressed as \cite{zhang2023calculation}:
\begin{equation}
E_{\mathrm{CO}_2} = \sum_i (E_i \cdot EF_i) \label{environment}
\vspace{-0.3cm}
\end{equation}
In \eqref{environment}, $E_{\mathrm{CO}_2}$ stands for the total port carbon emissions (kg~CO$_2$), $E_i$ is the consumption of energy $i$ (usually fossil fuels in this equation), and $EF_i$ is the carbon emission factor for the corresponding type of energy (kg~CO$_2$/kWh or kg~CO$_2$/L). To reduce carbon emissions, RES is adopted and the updated value of $E_{\mathrm{CO}_2}$ is calculated using \eqref{upatedeq} from \cite{jiangsu_ecology_2024} and \eqref{updated_1} computes the total reduction in  carbon emissions.
\begin{align}
    E_{\mathrm{CO}_2}^{\mathrm{new}} = \sum_i (E_i^{\mathrm{new}} \cdot EF_i) - E_{\mathrm{renew}} \cdot E_{\mathrm{CO}_2}^{\mathrm{grid}} 
\label{upatedeq} 
\end{align}

Where, $E^{\mathrm{total}}_{\mathrm{CO}_2} = E_{\mathrm{CO}_2} - E_{\mathrm{CO}_2}^{\mathrm{new}} \label{updated_1}$ and  in equation \eqref{upatedeq}, $E_{\mathrm{renew}}$ stands for the renewable energy supply (kWh) and $E_{\mathrm{CO}_2}^{\mathrm{grid}}$ is the grid average carbon emission factor (kg~CO$_2$/kWh). Here our objective is to minimize the total carbon emissions  $E^{\mathrm{total}}_{\mathrm{CO}_2}$.

\subsubsection{Energy Consumption Model}
It is assumed that energy consumption within a harbor can be categorized into three primary components: energy utilized by heavy machinery, which includes quay cranes and rubber-tire gantry cranes; energy expended by the transportation sector, encompassing trailers and cargo transport; and energy derived from the flexible load associated with building operations. Consequently, the overall energy consumption in the harbor is the summation of these distinct elements.  The total energy consumption of the harbour can be calculted as: $E^{\mathrm{total}}_{\mathrm{energy}} = E_{\mathrm{equip}} + E_{\mathrm{transport}} + E_{\mathrm{buildings}}$

In this equation, $E_{\mathrm{equip}}$ is the energy consumption of equipment such as cranes and automated guided vehicles (AGVs), $E_{\mathrm{transport}}$ stands for the energy consumption of trucks and container transport, and $E_{\mathrm{buildings}}$ is the energy consumption for port building lighting, air conditioning, etc.  Here our objective is to optimize the energy consumption  $E^{\mathrm{total}}_{\mathrm{energy}}$.

\subsubsection{Intelligent Transport System Dispatch}
In the context of a harbor that incorporates an automated guided vehicle (AGV) system for transportation tasks, the precise determination of optimal distances for the loading and unloading of goods is crucial. Consequently, here we aim  to minimize transportation distances while simultaneously enhancing operational efficiency.
\vspace{-0.3cm}
\begin{equation}
\min \sum_i \sum_j d_{ij} x_{ij}, \sum_j x_{ij} = 1, \quad \sum_i x_{ij} = 1
\end{equation}
Where $d_{ij}$ is the distance from point $i$ to point $j$, and $x_{ij}$ equals 1 if the AGV takes that path, 0 otherwise.
\vspace{-0.2cm}
\subsection{Solar energy }
The Yangshan Phase IV Terminal integrates an 8.5~MW photovoltaic system, generating approximately 8~million~kWh annually---meeting 30\% of its energy needs. This solar output powers automated equipment and shore power systems, coordinated via a smart EMS. The integration enhances energy efficiency while reducing carbon emissions.
The output power of photovoltaic power generation can be expressed as $P = A \cdot G \cdot \eta \label{S_1}$ and total annual power generation is given by \eqref{s_2} :
\begin{align}
    E_{\text{year}}^{\text{PV}} = P_{\text{peak}} \cdot H_{\text{sun}} \cdot PR \label{s_2}
\end{align}
where $P$, $A$, and $G$ represent the instantaneous output power, the total area of PV panels, and the solar irradiance, and $\eta$ represents the efficiency of the PV module (typically about 15\% to 20\%). Moreover, in \eqref{s_2}
$P_{\text{peak}}$ represents the peak power of the system, \(H_{\text{sun}}\) represents the average effective sunshine hours of the year, and \(PR\) is the performance ratio.

\subsection{Wind energy}
Though wind resources are moderate, small-scale wind turbines have been installed as supplemental sources. These systems, managed through the EMS, contribute during low solar output periods. While wind energy plays a minor role, it adds resilience to the port’s renewable energy mix and supports microgrid stability.
Similar to solar, we first calculate the instantaneous power of wind power:

\begin{align}
P_{\text{wind}} = \frac{1}{2} \cdot \rho \cdot A \cdot v^3 \cdot C_p \label{w_1} & 
\end{align}

Where, $
E_{\text{year}}^{\text{Wind}} = P_{\text{avg}} \cdot T_{\text{op}}$ and in  equation \eqref{w_1}, \(\rho\) represents the air density and is taken as a standard value of \(1.225 \, \text{kg/m}^3\), \(A\) is the swept area of the wind turbine, while \(v\) and \(C_p\) are the wind speed, and  power coefficient, which is typically taken as 0.35–0.45 respectively. In \eqref{w_2}\(P_{\text{avg}}\) represents the average annual power generation and \(T_{\text{op}}\) is the annual operating time.

\section{Methods}
\subsection{Hungarian Algorithm}
The Hungarian algorithm is a classical combinatorial optimisation algorithm for solving optimal assignment problems, where the objective is to find a minimum (or maximum) total cost assignment of tasks in a cost matrix such that each ‘task’ is uniquely assigned to a ‘worker’ (or resource), and each worker can only perform one task. In this model it is used for path scheduling of AGVs. In logistics transport, each AGV needs to move from a certain starting point to a target point, and the total path length of all AGVs is required to be the shortest, and at the same time, each AGV can only match with one target point without repetition.
\vspace{-0.3cm}
\begin{align}
    Cost_{Matrix}= 
\begin{bmatrix}
420 & 350 & 450 \\
450 & 400 & 280 \\
420 & 360 & 390
\end{bmatrix}
\end{align}

This matrix represents the path distances of the 3 AGVs from the starting points to the destinations A, B, and C. Assuming this matrix $C = c_{ij}$, where $c_{ij}$ denotes the assignment of the ith task to the jth AGV, the final total distance is
\vspace{-0.3cm}
\begin{align}
    Cost_{Matrix} = \sum_{i=1}^{3} \text{cost\_matrix}[i, \pi(i)]
\end{align}

\section{Problem formulation}
Building upon the established proposed solutions, this section formulates the optimization problem for the EMS integrator. Our objective is to develop a strategy that simultaneously addresses operational goals while leveraging renewable energy sources. Specifically, the problem formulation aims to achieve the following: Carbon emission minimization ( $E_{\mathrm{CO}_2}^{\mathrm{new}}$): Reduce the overall carbon footprint of port operations, energy Consumption minimization ($E^{\mathrm{total}}_{\mathrm{energy}}$): optimize energy usage to decrease total consumption within the port, AGV distance minimization ($c_{i,\pi(i)}$): minimize the travel distance for AGVs to enhance operational efficiency. Wind and PV energy maximization ($E_{res}$): maximize the utilization of available wind and solar power generation within the port's energy mix.

This multi-objective optimization problem is formulated below:

\begin{align}
   \min \sum_{i=1}^{N} \left( E_{\mathrm{CO}_2, {(i)}}^{\mathrm{new}} + E^{\mathrm{total}}_{\mathrm{energy},  {(i)}} + c_{i,\pi(i)}+ E_{res,(i)}\right) 
\end{align}


\section{Results and dicussion}

\begin{figure}[h]
    \centering
    \includegraphics[width=0.35\textwidth]{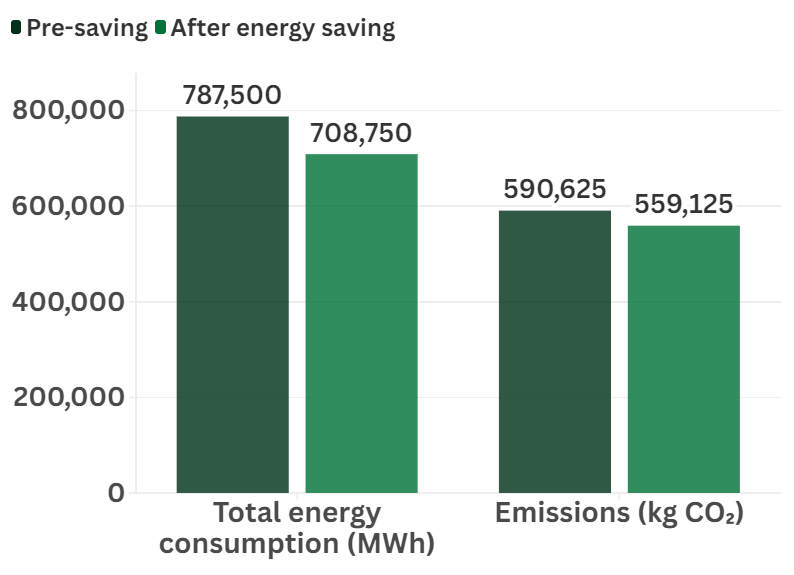}
    \caption{Total energy consumption comparison} \label{enrgy_2}
        \label{cyc_2}
    \vspace{-0.3cm}
\end{figure}

In this section, we present the simulation results derived from metrics related to energy consumption, ${\mathrm{CO}_2}$ emissions, AGV route optimization, and cost analysis. The simulations were conducted utilizing data from the design simulation environment of the Yangshan Phase IV terminal in Shanghai, implemented in Python. As shown in  Figure \ref{cyc_2}
baseline energy consumption is computed based on a throughput of 6.3 million TEUs and a unit energy consumption of 125 kWh/TEU, resulting in a total of 787,500 MWh. This consumption is allocated among equipment (50\%), transportation (30\%), and buildings (20\%). Applying emission factors of 0.5, 0.7, and 1.2 kg CO$_2$/MWh respectively, the initial CO$_2$ emissions amount to 590,625 kg CO$_2$. Incorporating 10\% renewable energy usage (78,750 MWh) with a grid emission factor of 0.4 kg CO$_2$/MWh reduces emissions by 31,500 kg CO$_2$, highlighting the environmental benefits of renewable integration and as well as saving energy consumption to  708,750 MWh per year.

It is worth noting that energy optimization reduced total consumption by 10\% (from 787,500 MWh to 708,750 MWh), but carbon emissions only decreased by 5.3\% (from 590,625 kg CO$_2$ to 559,125 kg CO$_2$), indicating a desynchronization between energy saving and emission reduction. Post-optimization, carbon intensity slightly rose from 0.75 kg CO$_2$/MWh to 0.79 kg CO$_2$/MWh, suggesting that measures prioritized lower-carbon intensity areas while high-carbon factors in building energy consumption (up to 1.2 kg CO$_2$/MWh) saw less reduction. Potentially, the system introduced 75,000 MWh of new green energy (solar and wind), achieving a significant 31,500 kg CO$_2$ reduction by replacing fossil with clean energy rather than reducing total consumption, particularly in hard-to-abate areas. This highlights that while energy conservation directly cuts costs, emission reduction, especially through green energy integration, often requires higher systemic investment, with a verified carbon substitution efficiency of 0.42 kg CO$_2$/MWh.

\begin{figure}[h]
    \centering
    \includegraphics[width=0.35\textwidth]{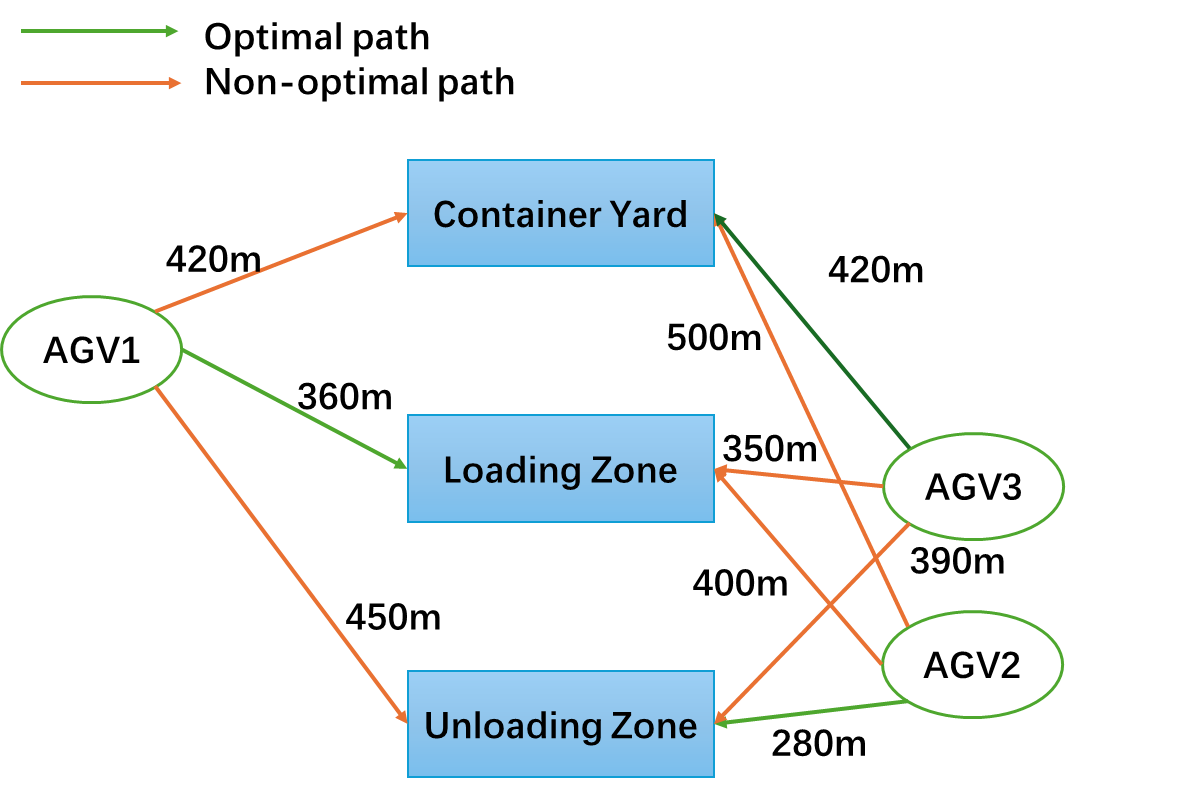}
    \caption{AGV Path Planning with the Hungarian Algorithm} \label{enrgy_3}
        \label{cyc_21}
    \vspace{-0.3cm}
\end{figure}

Morover, the optimization of AGV routes is conducted to attain the shortest possible trajectory for AGV operations. This optimization process involves the selection of designated nodes within the Shanghai Yangshan Phase IV, utilizing the Hungarian algorithm. The application of the Hungarian algorithm serves to enhance AGV routing between various nodes in significant port facilities. By employing a distance matrix that quantifies the distances between nodes (for instance, a distance of 35 kilometers between nodes A and B), the algorithm effectively identifies the optimal allocation that minimizes the total distance traversed as shown in \ref{enrgy_3}. Consequently, this optimization contributes to the enhancement of logistical efficiency within the port environment.
Finally, operating costs were assessed on the basis of a throughput of 6.3 million TEUs, with baseline and optimised costs set at \$250 and \$175 per TEU, respectively. This resulted in a total cost reduction of \$472.5 million, or 30 per cent. These figures highlight the significant economic advantages of implementing EMS optimisation in port operations.
\vspace{-0.3cm}

\begin{table}[h]
\centering
\caption{Cost Comparison Per TEU and in Total}
\begin{tabular}{lccc}
\toprule
\textbf{Item} & \textbf{Baseline} & \textbf{Optimized} & \textbf{Savings} \\
\midrule
Per TEU & \$250 & \$175 & \$75 \\
Total (6.3M TEU) & \$1,575M & \$1,102.5M & \$472.5M (30\%) \\
\bottomrule
\end{tabular}
\label{tab:cost_comparison}
\vspace{-0.5cm}
\end{table}

\section{Conclusion}
This paper examines the role of EMS in the context of smart ports, which can yield significant enhancements in operational efficiency, environmental sustainability, and cost-effectiveness. Initially, we conduct a comprehensive analysis of case studies from four representative smart ports: Hamburg Port, Genoa Port, Jurong Port, and Shanghai Yangshan Phase IV. This analysis is framed around several key parameters, including the average share of renewable energy, reductions in emissions, cost savings, and energy demand efficiency. Our study facilitates a systematic comparison among the selected ports and underscores the advanced capabilities of EMS, key driving factors, core technologies, and technical tools essential for the implementation of sustainable and cost-effective solutions in smart port operations. Furthermore, we develop an optimization model to simulate load dispatch, carbon emission reduction, and transportation scheduling. The results demonstrate that the deployment of EMS can lead to significant reductions in annual energy consumption and carbon emissions, estimated at approximately 7–8\% and 11–12\%, respectively, while simultaneously achieving considerable cost savings.
\section*{Acknowledgments}
This work is partially supported by the Research Council of Finland through ECO-NEWS n.358928 and X-SDEN n. 349965, and by EU MSCA project ``COALESCE'' under Grant Number 101130739.

\bibliographystyle{ieeetr}
\bibliography{Ref}
\end{document}